\documentclass[aip,amsmath,amssymb,graphicx,reprint]{revtex4-1}

\usepackage{graphicx}
\usepackage{cancel}
\usepackage{dcolumn}
\usepackage{xcolor}

\draft 

\begin{document}

\title{Rovibrational structure of the Ytterbium monohydroxide molecule and the $\mathcal{P}$,$\mathcal{T}$-violation searches}

\author{Anna Zakharova} \email{zakharova.annet@gmail.com}
\affiliation{St. Petersburg State University, St. Petersburg, 7/9 Universitetskaya nab., 199034, Russia} 
\affiliation{Petersburg Nuclear Physics Institute named by B.P. Konstantinov of National Research Centre
"Kurchatov Institute", Gatchina, 1, mkr. Orlova roshcha, 188300, Russia}
\author {Igor Kurchavov}\email{igorkurchavov58@gmail.com}
\affiliation{Petersburg Nuclear Physics Institute named by B.P. Konstantinov of National Research Centre
"Kurchatov Institute", Gatchina, 1, mkr. Orlova roshcha, 188300, Russia}
\author {Alexander Petrov}\email{petrov\_an@pnpi.nrcki.ru}

\affiliation{St. Petersburg State University, St. Petersburg, 7/9 Universitetskaya nab., 199034, Russia} 
\affiliation{Petersburg Nuclear Physics Institute named by B.P. Konstantinov of National Research Centre
"Kurchatov Institute", Gatchina, 1, mkr. Orlova roshcha, 188300, Russia}

\date{\today}

\begin{abstract}
{ The spectrum of  triatomic molecules with close rovibrational opposite parity levels is sensitive to the $\mathcal{P}$,$\mathcal{T}$-odd effects. This makes them a convenient platform for the experimental search of a new physics. Among the promising candidates one may distinguish the YbOH as a non-radioactive compound with a heavy atom. 
The energy gap between levels of opposite parity, $l$-doubling, is of a great interest as it determines the electric field strength required for the full polarization of the molecule. Likewise, the influence of the bending and stretching modes on the sensitivities to the $\mathcal{P}$,$\mathcal{T}$-violation requires a thorough investigation since the measurement would be performed on the excited vibrational states. This motivates us to obtain the rovibrational nuclear wavefunctions, taking into account the anharmonicity of the potential. As a result, we get the values of the $E_{\rm eff}$ and $E_s$ for the lowest excited vibrational state and determine the $l$-doubling
}
\end{abstract}


\maketitle 
\section{Introduction}
The violation of the charge conjugation ($\mathcal{C}$), spatial reflection ($\mathcal{P}$), and time reversal ($\mathcal{T}$) symmetries is the striking feature of the Standard model (SM) \cite{khriplovich2012cp,schwartz2014quantum,particle2020review}. The sources of the charge-parity ($\mathcal{CP}$) nonconservation in the SM are Cabibbo-Kobayashi-Maskawa (CKM) \cite{Cabibbo1963,KobayashiMaskawa1973} and Pontecorvo–Maki–Nakagawa–Sakata (PMNS) matrices \cite{Pontecorvo1957,MNS1962}, and, possibly, the $\theta$ term of the strong interaction \cite{Cheng1988,KimCarosi2010}. One of the $\mathcal{CPT}$ theorem consequences is that nonconservation of $\mathcal{CP}$ is equal to the violation of $\mathcal{T}$-symmetry.

One of the possible manifestation of the $\mathcal{CP}$-nonconservation is the electron electric dipole moment (eEDM). In the Standard model the eEDM appears only in the multiloop processes with a high order of the weak coupling constant and, thus, the predicted value is very small. On the other hand, some models of the physics beyond the Standard Model (SM) forecast new $\mathcal{CP}$-violation sources that can lead to the significant increase of the eEDM \cite{Fukuyama2012,PospelovRitz2014,YamaguchiYamanaka2020,YamaguchiYamanaka2021}. The presence of the particle superpartners in the Supersymmetry theory (SUSY) would provide new $\mathcal{P}$, $\mathcal{T}$-violation sources. Besides, the fluctuations of the $\theta$ parameter, the axion, in the Peccei–Quinn theory may result in the  $\mathcal{CP}$-violating processes. Furthermore, the matter-antimatter ratio in the observable universe  \cite{DineKusenko2003, Sakharov1967} may imply new sources of the charge-parity violation.

The high precision molecular experiments provide a powerful way to investigate the $\mathcal{CP}$-violating physics \cite{baron2014order,ACME:18}.

As for now, the best experimental bound on the eEDM was obtained for diatomic molecules with closely spaced $\Omega$-doublets such as ThO \cite{ACME:18,DeMille:2001,Petrov:14,Vutha:2010,Petrov:15,Petrov:17} and  HfF$^{+}$ \cite{Cornell:2017,Petrov:18}. These experiments also put constraints on the scalar-pseudoscalar nucleon-electron interaction \cite{ginges2004violations,PospelovRitz2014,ChubukovLabzowsky2016}.
Experiments for searching other $\mathcal{P}$, $\mathcal{T}$ odd effects, including nuclear magnetic quadrupole moment \cite{FDK14, Petrov:17b, Kurchavov:20, maison2019theoretical}  and axion mediated interactions \cite{maison2020study, maison2021axion} are planed.

The vibrational modes of the polyatomic molecules create unique spectral characteristics not possessed by diatomic molecules. For instance, the triatomic species can simultaneously allow  laser-cooling \cite{Isaev_2017} and possess levels with opposite parity, the so-called $l$-doublets \cite{Kozyryev:17,hutzler2020polyatomic}.

The levels of opposite parities constituting the $l$-doublet are mixed when the external electric field applied so that the molecule becomes polarized. The $\mathcal{P}$, $\mathcal{T}$-violation is manifested in the energy splitting, $\Delta E_{\mathcal{P},\mathcal{T}}$, between  the levels with opposite values $\pm M$ of  total angular momentum projection on the electric field axis.

If the electron has EDM $d_e$ and is affected by the scalar-pseudoscalar interaction with nuclei characterized by the coupling constant $k_s$, these $\mathcal{P}$, $\mathcal{T}$-odd effects can be estimated from the  maximum splitting between levels with opposite values of $M$ given by,
\begin{equation}
\Delta E_{\mathcal{P},\mathcal{T}}=2E_{\rm eff}  d_e + 2E_{\rm s} k_s,
\label{split}
\end{equation}
The parameters $E_{\rm eff}$ and $E_{\rm s}$ are determined by the molecular electronic structure \cite{KozlovLabzowsky1995, titov2006d, Safronova2017}.

The laser-cooling in one dimension was achieved for alkaline earth metal monohydroxides such as SrOH \cite{kozyryev2017sisyphus} and, recently, the YbOH \cite{steimle2019field,augenbraun2020laser}. The latter is considered a promising candidate for the future experiments searching eEDM \cite{Kozyryev:17}.

The sensitivity of the 
YbOH molecule to the eEDM was previously computed in \cite{denis2019enhancement} within the relativistic coupled cluster method. However, the vibrational motion, including the bending modes, of excited vibrational states may influence the value of this parameter. In \cite{prasannaa2019enhanced} $E_\mathrm{eff}$ was studied for different nonlinear configurations and strong dependence on the bending
angle already at Dirac-Hartree-Fock (DHF) level was stressed. This claim is inconsistent with the results of \cite{gaul2020ab} that used the complex generalized Hartree-Fock and Kohn-Sham methods within zeroth-order regular approximation and also has given the harmonic estimate for $E_\mathrm{eff}$ at the $v=1$ vibrational level.

Previously we obtained the rovibrational
wavefunctions for the molecule RaOH \cite{ourRaOH}. This allowed us not only to compute the $E_\mathrm{eff}$ and $E_\mathrm{s}$ parameters for the first vibrational levels but also to obtain the value of the $l$-doubling that determines the external electric field required for the complete polarization of the molecule. In this paper, we apply the techniques we developed to perform a similar analysis for the YbOH molecule.

\section{Methods}

\begin{figure}[h]
\centering
  \includegraphics[width=0.25\textwidth]{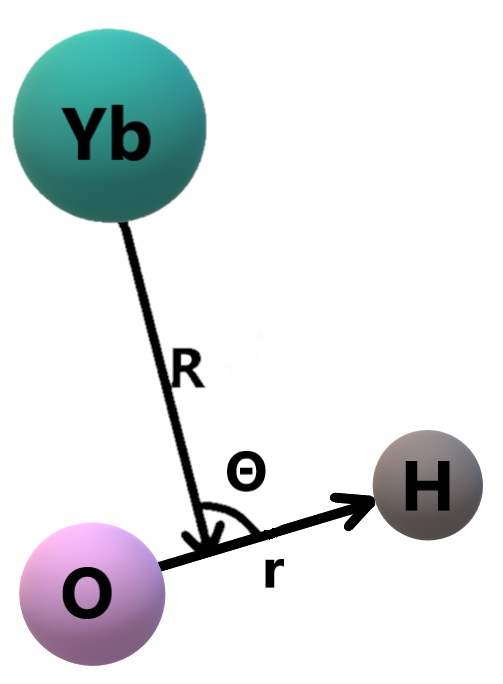}
  \caption{Jacobi coordinates}
  \label{Jacob}
\end{figure}

We assume that the wavefunction of the molecule can approximately be factorized into the nuclear and electronic parts,
\begin{equation}
\Psi_{\rm total}\simeq\Psi_{\rm nuc}(Q)\psi_{\rm elec}(Q|q),
\label{totalWF}
\end{equation}
where $Q$ denotes generalized coordinates of the nuclei and $q$ - generalized coordinates of the electrons. Within the Born-Oppenheimer approximation the electronic part $\psi_{\rm elec}(Q|q)$ is the solution of the Dirac-Coulomb equation for the electrons in the field of the nuclei fixed at coordinates $Q$. To describe the configuration of the triatomic molecule we choose $Q$ as the Jacobi coordinates represented in Fig.~\ref{Jacob}: $\hat{r}$ and $\hat{R}$ are unit 
vectors directed along the OH axis and Yb - OH center of mass (c.m.) axis respectively, $\theta$ is the angle between above axes, $R$ is the distance between Yb and the c.m. of OH. As the frequency of OH vibrational mode is about one order of magnitude larger than other vibrational frequencies in YbOH, we fix OH ligand stretch at the equilibrium distance $r=1.832 \,a.u.$ \cite{Huber:1979}.

The nuclear part of the wavefunction $\Psi_{\rm nuc}$ satisfies the Schr\"{o}dinger equation,
\begin{equation}
\hat{H}_{\rm nuc}\Psi_{\rm nuc}(R, \hat{R}, \hat{r}) = E \Psi_{\rm nuc}(R, \hat{R}, \hat{r}).
\label{Shreq}
\end{equation}
The 
nuclear
Hamiltonian takes the form,
\begin{equation}
\hat{H}_{\rm nuc}=-\frac{1}{2\mu}\frac{\partial^2}{\partial R^2}+\frac{\hat{L}^2}{2\mu R^2}+\frac{\hat{j}^2}{2\mu_{OH}r^2}+V(R,\theta),
\end{equation}
where $\mu$ is the the $Yb-OH$ reduced mass, $\mu_{OH}$ is the the OH ligand reduced mass, $\hat{L}$  is the angular momentum of the Yb-OH system rotation 
around its c.m., $\hat{j}$ is the ligand angular momentum, and $V(R,\theta)$ is the effective adiabatic potential obtained from the electronic structure calculations.

The sensitivity of the spectrum to the $\mathcal{P}$, $\mathcal{T}$-odd interactions for the fixed configurations can be described by the parameters,
\begin{equation}
E_{\rm eff}(R,\theta)=\frac{\langle\psi_{\rm elec}(R,\theta)| \hat{H}_d|\psi_{\rm elec}(R,\theta)\rangle}{d_e{\rm sign}(\Omega)},
\end{equation}
\begin{equation}
E_{\rm s}(R,\theta)=\frac{\langle\psi_{\rm elec}(R,\theta)| \hat{H}_s|\psi_{\rm elec}(R,\theta)\rangle}{k_s{\rm sign}(\Omega)},
\end{equation}
that can be understood as the expectation values for the eEDM and scalar-pseudoscalar nucleon-electron interaction terms in the $\mathcal{P}$, $\mathcal{T}$-odd interaction Hamiltonian Hamiltonian
\begin{equation}
\hat{H}_{\cancel{\mathcal{PT}}}=\hat{H}_d+\hat{H}_s,
\end{equation}
\begin{equation}
\hat{H_d}=  2d_e\sum_{i}
  \left(\begin{array}{cc}
  0 & 0 \\
  0 & \bf{\sigma_i E_i} \\
  \end{array}\right)\ 
 \label{Hd},
\end{equation}
\begin{equation}
\hat{H_s}=ik_s\frac{G_F}{\sqrt2}\sum_{j=1}^{N_{\rm elec}}\sum_{I=1}^{N_{\rm nuc}}{\rho_I\left(\vec{r_j}\right)Z_I}\gamma^0\gamma^5,
\label{Hs}
\end{equation}
where  $\rho_I$ is the normalized charge density of the  $I$-th nucleon, $G_F$ is Fermi constant, $\bf{\sigma}$ are Pauli matrices, $\bf{E_i}$ is the internal molecular electric field that acts on ith electron.

For the total molecular wavefunction (\ref{totalWF}) these parameters should be averaged over the nuclear wavefunction,
\begin{equation}
\label{Eeffaver}
E_{\rm eff,s}=\int dR d\hat{R} d\hat{r} |\Psi_{\rm nuc}(R, \hat{R}, \hat{r})|^2 E_{\rm eff,s}(R,\theta).
\end{equation}

The Ytterbium atom was described by a 28-electron generalized relativistic effective core potential (GRECP) \cite{titov1999generalized,mosyagin2010shape,mosyagin2016generalized} and a 42-valence electron basis set developed by the PNPI Quantum Chemistry Laboratory \cite{QCPNPI:Basis}. The cc-pVTZ basis was used for H and O atoms. The calculations were performed on a grid of Jacobi coordinates. The $R$ coordinate ranges from $2.6\, a.u.$ to $4.3\,a.u.$ with step $0.1\, a.u.$ The $\theta$ angle values are $0^\circ$, $5^\circ$, $10^\circ$, $15^\circ$, $20^\circ$, $25^\circ$, $57^\circ$, $90^\circ$, $122^\circ$, $155^\circ$ and $180^\circ$. Extra points near the equilibrium were added to better describe the region most relevant for the lowest vibrational levels.

The molecular two-component pseudospinors were obtained using the Hartree-Fock self-consistent field (SCF) method implemented in the Dirac 19 software \cite{DIRAC19}. The pseudospinors
are smoothed in the inner core region, so that the electronic
density in this region is not correct. The operators
in eqs. (\ref{Hd},\ref{Hs}) are heavily concentrated near the
nucleus and are therefore strongly affected by the wave
function in the inner region. The four-component molecular
spinors must therefore be restored in the inner region
of Yb.
The MOLGEP program was used to apply the method of one-center restoration of the correct four-component spinors in the core region with help of the equivalent basis sets \cite{Petrov:02,titov2006d,skripnikov2015theoretical}. The matrix elements of $\hat{H}_\mathrm{d}$ and $\hat{H}_\mathrm{s}$ were computed in the basis of the restored spinors $\psi_i$.

Restoration of the basis begins with the creation of an equivalent basis set of atomic
(one-center) four-component spinors:
  \begin{equation}
  \left\{\left( \begin{array}{c}
  f_{nlj}(r)\chi_{ljm} \\
  g_{nlj}(r)\chi_{l'jm}
  \end{array} \right) \right\},
  \end{equation}
and two-component pseudospinors $\left\{ \tilde{f}_{nlj}(r)\chi_{ljm}\right\}$. Here $f$ - large component, $g$ - small
component, $\chi$ - spin-angular part, $n$ - principal quantum number, $j$ and $m$ - total
electronic moment and his projection in internuclear axis, $l$ and $l'$ - orbital moment, and $l'=2j-l$.

For the numerical four-component and two-component atom calculations, the HFD and
HFJ/GRECP programs were used to create two equivalent basis sets for
reconstruction. Molecular pseudo-orbitals then decompose in the basis of
two-component single-center atomic pseudospinors,

\begin{equation}
  \tilde{\phi}_i(\mathrm{r})\approx \sum_{l=0}^{L_{\text{max}}}\sum_{j=|l-1/2|}^{j=|l+1/2|}\sum_{nm}c_{nljm}^i\tilde{f}_{nlj}(\mathrm{r})\chi_{ljm}.
\end{equation}
Then two-component pseudospinors are replaced by equivalent four-component spinors:
\begin{equation}
  \label{eq4}
  \phi_i(\mathrm{r})\approx \sum_{l=0}^{L_{\text{max}}}\sum_{j=|l-1/2|}^{j=|l+1/2|}\sum_{nm}c_{nljm}^i\left( \begin{array}{c}
  f_{nlj}(r)\chi_{ljm} \\
  g_{nlj}(r)\chi_{l'jm}
\end{array} \right).
\end{equation}
Molecular four-component spinors constructed in this way are orthogonal to the core
spinor Yb, since atomic basis functions in the equation were calculated for frozen inner
core electrons.
For the current calculation we put $L_{\text{max}}=3$, what is enough for accurate calculation of $E_{\rm eff}$ and $E_\mathrm{s}$ \cite{Petrov:02}.

\begin{figure}[h]
\centering
  \includegraphics[width=0.5\textwidth]{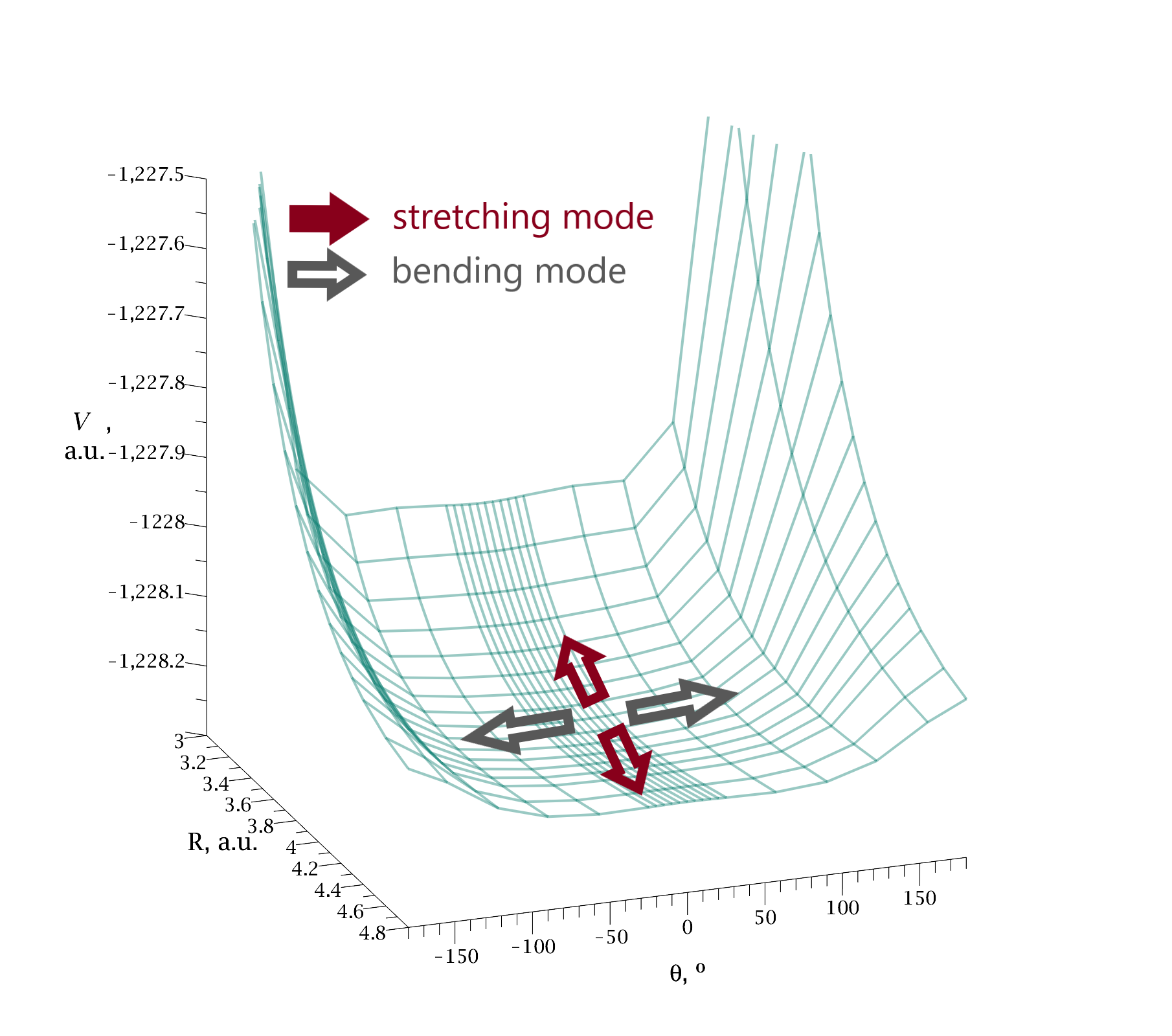}
  \caption{Potential surface $V(R,\theta)$}
  \label{Surface}
\end{figure}

For the correlation computations we have chosen the active space with 30 frozen electrons and 21 active ones. The relativistic coupled cluster method with single, double and perturbative triple excitations (CCSD (T)) implemented in Dirac 19 RELCCSD module was used to compute the points of the potential surface on the grid defined above. Then the function $V(R,\theta)$ was constructed by the bicubic interpolation Fig.~\ref{Surface}.

The obtained adiabatic potential was used for numerical computations of the nuclear wavefunctions based on the close-coupled equations \cite{mcguire1974quantum} obtained as a decomposition of (\ref{Shreq}) in terms of the eigenfunctions of the molecular and ligand angular momenta. For the details of our approach we refer the reader to our previous work \cite{ourRaOH}. This way we take into account the anharmonicities of the potential and interaction between the rotational and vibrational degrees of freedom without resorting to the perturbative techniques.

\begin{figure*}[h!]
\centering
  \includegraphics[width=0.49\textwidth]{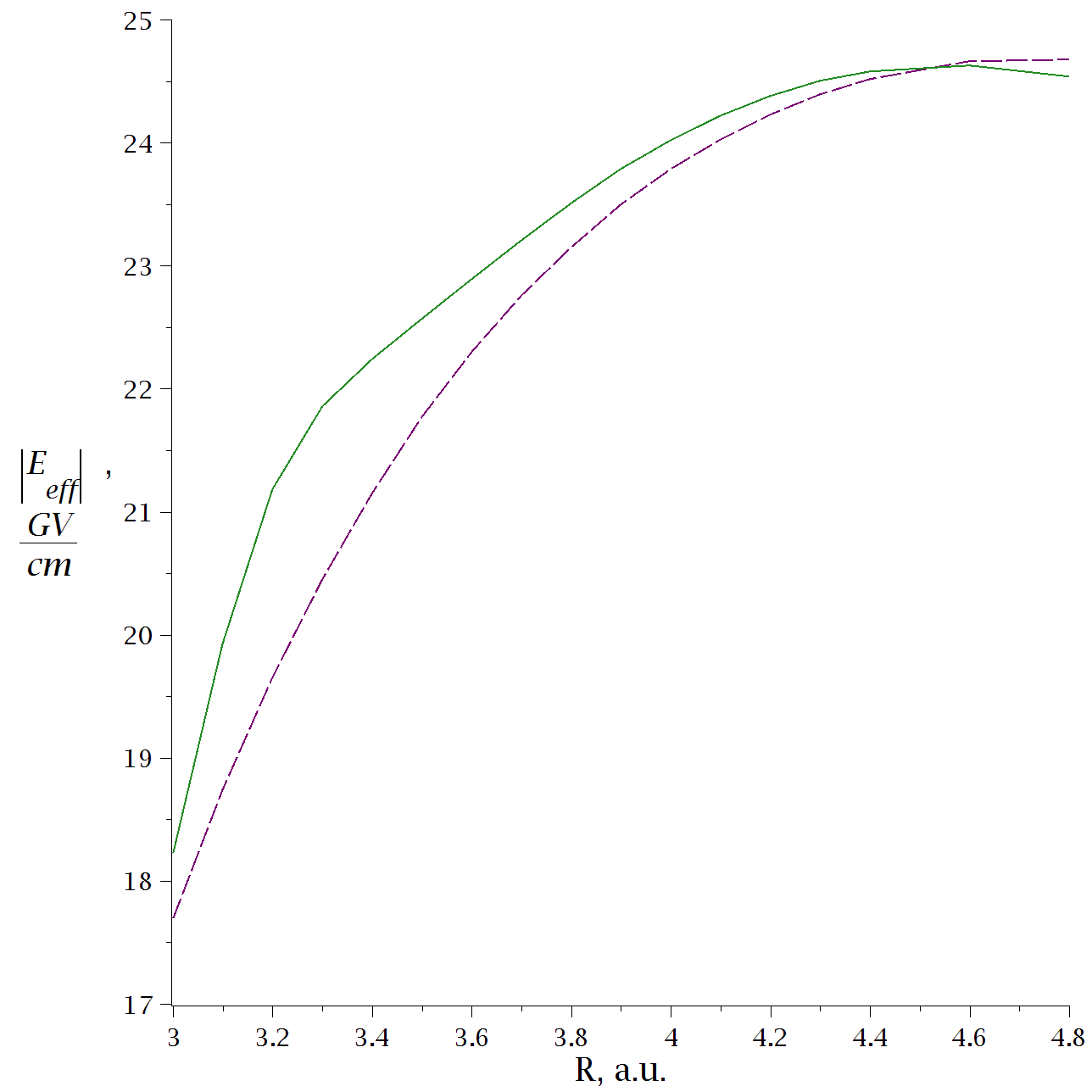}
  \includegraphics[width=0.49\textwidth]{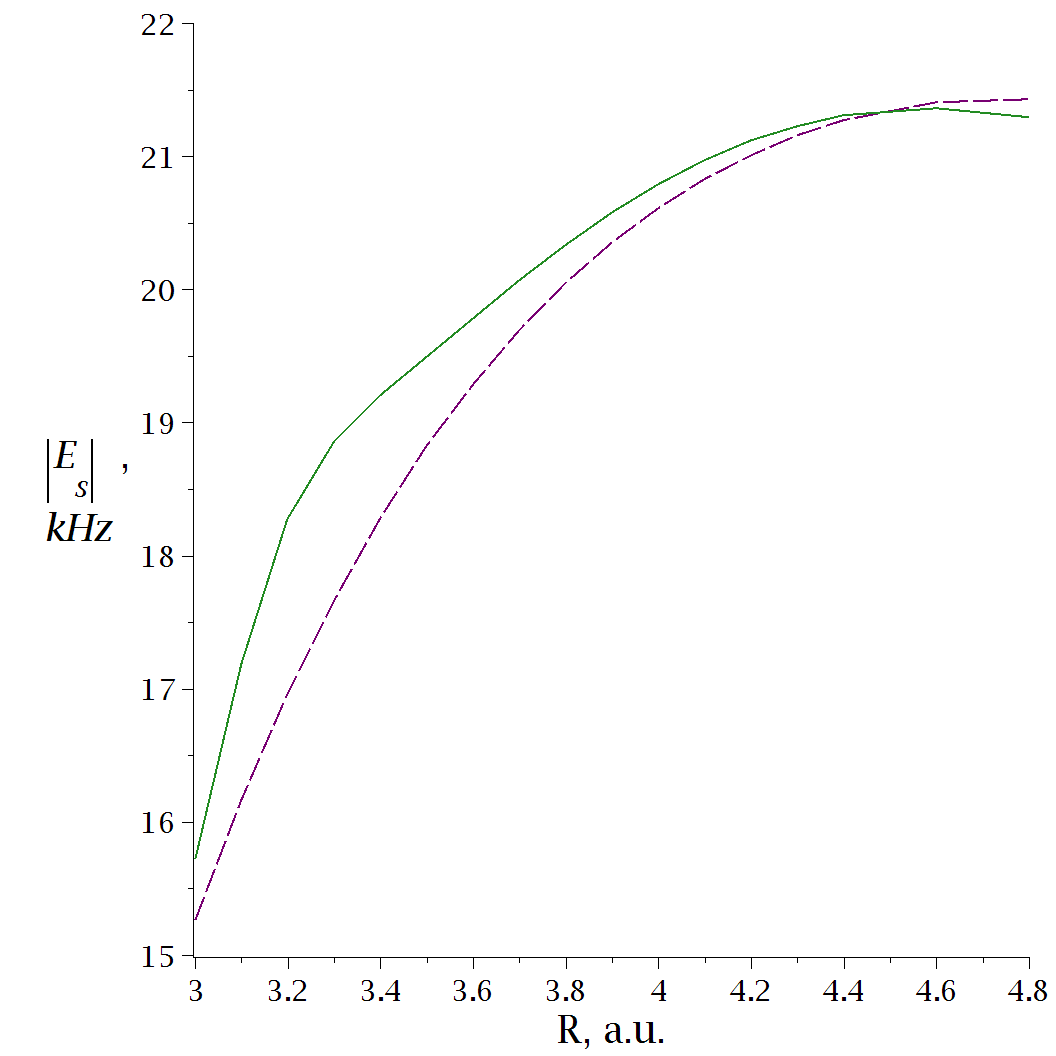}
  \caption{SCF (dashed) and CCSD (solid) $\mathcal{P}$,$\mathcal{T}$-odd parameters for linear configurations of YbOH}
  \label{PTLinear}
\end{figure*}

For the property calculations of the linear configuration we get the one-electron density matrix $\rho^{(1)}_{ij}$ at CCSD level with help of the MRCC program suite \cite{MRCC2020}. The density matrix is then contracted with $E_{\rm eff}$ and $E_{\rm s}$ matrix elements to obtain the values of the correlation
corrections for the linear configurations,
\begin{equation}
E_{\rm eff,s}(R,\theta)=\frac{1}{N_\mathrm{{elec}}}\sum_{i,j=1}^{N_{\rm orb}}\rho^{(1)}_{ij}\frac{\langle \psi_i|\hat{H}_{\rm d,s}|\psi_j\rangle}{d_e{\rm sign}(\Omega)}.
\end{equation}
Regretfully, the Dirac-MRCC interface works only for the symmetry groups with real representations (such as $C_{2v}$) but not with  ones with complex representations (such as $C_s$) of the nonlinear molecules. Therefore the CCSD correction was obtained only for the linear configurations and are depicted on Fig. \ref{PTLinear}. Since these corrections constitute only about $1\%$ of the SCF values near minimum reaching $6\%$ far from equilibrium point, we assumed that, as a first approximation, it is reasonable to approximate the CCSD correction for the nonlinear molecule by the result computed for the linear molecule:
\begin{align}
&E_{{\rm eff},{\rm s}}^{({\rm total},I)}(R,\theta)=E_{{\rm eff},{\rm s}}^{({\rm ccsd})}(R,0^\circ)+ \nonumber\\
&+\Big( E_{{\rm eff},{\rm s}}^{({\rm scf})}(R,\theta)- E_{{\rm eff},{\rm s}}^{({\rm scf})}(R,0^\circ)\Big).\label{EtotalI}
\end{align}

In other words, the dependence of $E_{\rm eff,s}(R,\theta)$ on $\theta$ was calculated at SCF level. To test the validity of our approximation, we made a finite field computation of
$E_{\rm eff,s}$
by CCSD method for the near equilibrium value of $R=3.9\,\mathrm{a.u.}$ and different angles. Because this computation is very expensive we were able to obtain the values only for the single value of $R$.

\section{Results and discussion}

The spectrum of the nuclear wavefunctions $\Psi_{\rm nuc}$ is characterized by the parameters we present in the Table~\ref{tab:Spectrum}. 
The agreement between the computed and the experimental values of frequencies is rather good.

Our $l$-doubling value is consistent with an estimate\cite{HerzbergBook},
\begin{equation}
q \simeq 
    \frac{B^2}{\nu_2}\Big(1+4\frac{\zeta_{21}^2\nu_{2}^2}{\nu_{1}^2-\nu_{2}^2}\Big)(v+1).
\end{equation} Comparing with our results, we can find Coriolis coefficient, $\zeta_{21}=0.265$. The value of $l$-doubling for YbOH is greater than our result for RaOH molecule, $\Delta E_{J=1}=2q=14.5$ \rm MHz\cite{ourRaOH} as expected from the smaller momentum of inertia  of the  YbOH molecule.

\begin{table}
\caption{\label{tab:Spectrum} Rovibrational spectrum parameters}
\begin{ruledtabular}
\begin{tabular}{ccc}
& Computation & Experiment\footnote{The number in parenthesis denotes $2\sigma$ deviation} \\
\hline
Stretching mode $\nu_1$
  & $550 {\rm cm}^{-1}$ & $529.341(1) {\rm cm}^{-1}$,\cite{melville2001visible} \\
Bending mode $\nu_2$
  & $319 {\rm cm}^{-1}$ & $339(5) {\rm cm}^{-1},$\cite{melville2001visible} \\
Rotational constant $B$  & $0.2461 {\rm cm}^{-1}$ & $0.245434(13) {\rm cm}^{-1}$,\cite{melville2001visible}\\
&& $0.2451163(10) {\rm cm}^{-1}$,\cite{nakhate2019pure}\\
$l$-doubling $\Delta E_{J=1}=2q$  & $26 {\rm MHz}$ & \\
\end{tabular}
\end{ruledtabular}
\end{table}

\begin{figure}[h!]
  \includegraphics[width=0.49\textwidth]{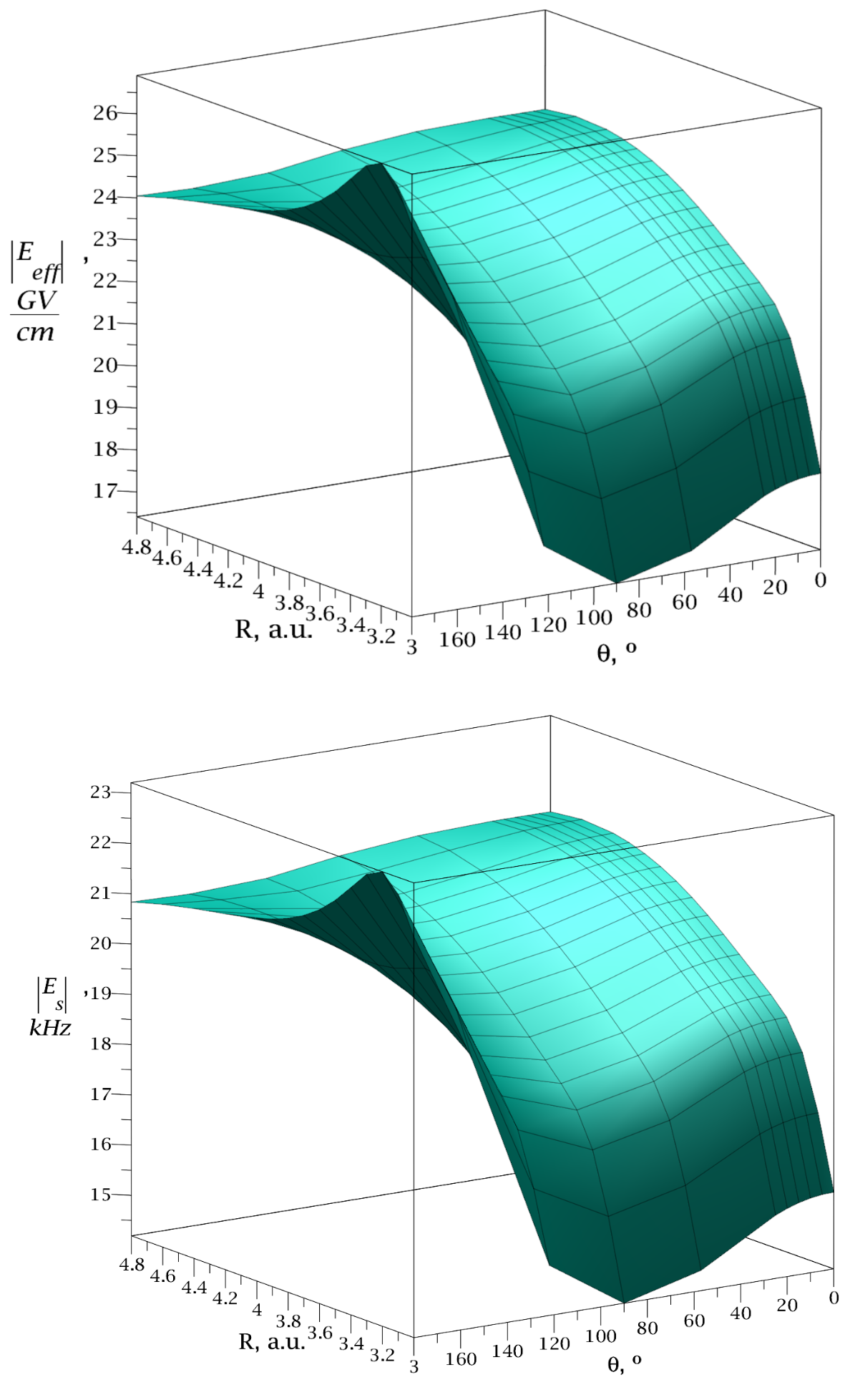}
  \caption{SCF $\mathcal{P}$,$\mathcal{T}$-odd parameters for
nonlinear configurations of YbOH}
  \label{PT3d}
\end{figure}

The dependence of the $E_{{\rm eff},{\rm s}}$ on the bending and stretching is depicted on the Fig.~\ref{PT3d}. As with RaOH \cite{ourRaOH} we do not confirm the oscillatory behavior claimed in \cite{prasannaa2019enhanced}.

As described in previous section, the full dependence of $E_{\rm eff,s}(R,\theta)$ on $\theta$ was calculated at SCF level. The finite field computation was performed only for the single value of $R=3.9\,\mathrm{a.u.}$. In the Table~\ref{tab:SCF_FF_Compare} we compare the deviations from the equilibrium value of $E_{\rm eff}(\theta)-E_{\rm eff}(0^\circ)$ for the SCF and correlation corrections results. The changes of the correlation correction happen to be of the same magnitude as the changes of the SCF values. The deviation becomes significant for large bending angles and, while this does not affect the average values on the rovibrational levels considered, it may become important for the higher excited levels. The similar analysis was made for the $E_{\rm s}$ values in the Table \ref{tab:SCF_FF_Compare2}. Unlike SCF values the correlation correction for $E_{\rm s}$ have a different angular dependence from $E_{\rm eff}$.

To take the angular dependence of the correlation correction $\Delta E^{(corr)}_{{\rm eff},{\rm s}}$ into account using the available data we use the following approximation,
\begin{align}
&E_{{\rm eff},{\rm s}}^{({\rm total},II)}(R,\theta)=E_{{\rm eff},{\rm s}}^{({\rm ccsd})}(R,0^\circ)+ \nonumber\\
&+\Big( E_{{\rm eff},{\rm s}}^{({\rm scf})}(R,\theta)- E_{{\rm eff},{\rm s}}^{({\rm scf})}(R,0^\circ)\Big)
\nonumber\\
&+\Big( E_{{\rm eff},{\rm s}}^{({\rm ccsd})}(3.9\,\mathrm{a.u.},\theta)- E_{{\rm eff},{\rm s}}^{({\rm scf })}(3.9\,\mathrm{a.u.},\theta)\Big)
\nonumber\\
&-\Big( E_{{\rm eff},{\rm s}}^{({\rm ccsd})}(3.9\,\mathrm{a.u.},0^\circ)- E_{{\rm eff},{\rm s}}^{({\rm scf })}(3.9\,\mathrm{a.u.},0^\circ)\Big).
\label{Etotal}
\end{align}

\begin{table}
\caption{\label{tab:SCF_FF_Compare} The deviations of $E_{\rm eff}$ for $R=3.9\,\mathrm{a.u.}$ from the equilibrium values in the finite field approach}
\begin{ruledtabular}
\begin{tabular}{ccc}
Angle & SCF, GV/cm & CCSD correction, GV/cm \\
\hline
$5^\circ$ & -0.006 & \,\,0.003 \\
$10^\circ$ & -0.025 & -0.024 \\
$15^\circ$ & -0.055 & -0.066 \\
$20^\circ$ & -0.096 & -0.117\\
$25^\circ$ & -0.149 & -0.173\\
$57^\circ$ & -0.629 & -0.209\\
$90^\circ$ & -0.923 & -0.722\\
$122^\circ$ & -0.773 & -5.138
\end{tabular}
\end{ruledtabular}
\end{table}

\begin{table}
\caption{\label{tab:SCF_FF_Compare2} The deviations of $E_{\rm s}$ for $R=3.9\,\mathrm{a.u.}$ from the equilibrium values in the finite field approach}
\begin{ruledtabular}
\begin{tabular}{ccc}
Angle & SCF, GV/cm & CCSD correction, GV/cm \\
\hline
$5^\circ$ & -0.006 & \,\,0.001 \\
$10^\circ$ & -0.022 & \,\,0.002 \\
$15^\circ$ & -0.049 & \,\,0.005 \\
$20^\circ$ & -0.086 & \,\,0.008\\
$25^\circ$ & -0.132 & \,\,0.010\\
$57^\circ$ & -0.557 & -0.030\\
$90^\circ$ & -0.816 & -0.297\\
$122^\circ$ & -0.685 & -0.658
\end{tabular}
\end{ruledtabular}
\end{table}

We summarize the results for the $\mathcal{P}$, $\mathcal{T}$-odd parameters and confront them with the preceding work in the Table~\ref{tab:EeffEs}. The sensitivities to $\mathcal{P}$,$\mathcal{T}$-odd effects is more than two times smaller than for the RaOH molecule\cite{ourRaOH}. Our results are in concordance with the previous estimates for the fixed geometries. While for the lower levels vibrations do not strongly affect the $E_\mathrm{eff}$ and $E_\mathrm{s}$ parameters, for higher levels it may become significant.

\begin{table}
\caption{\label{tab:EeffEs} Sensitivities to the $\mathcal{P}$, $\mathcal{T}$-odd effects for YbOH}
\begin{ruledtabular}
\begin{tabular}{ccc}
& $E_{\rm eff},\,\mathrm{GV}/\mathrm{cm}$ & $E_{\rm s}, kHz$ \\
\hline
Equilibrium geometry
  & $23.875$ & $20.659$ \\
 \hline
\multicolumn{3}{l}{Angular dependence as in (\ref{EtotalI})}\\
$v=0$ state
  & $23.810$ & $20.602$ \\
$v=1$ state
  & $23.740$ & $20.540$ \\
\hline
\multicolumn{3}{l}{Angular dependence  as in (\ref{Etotal})}\\
$v=0$ state
  & $23.716$ & $20.608$ \\
$v=1$ state
  & $23.576$ & $20.548$ \\
\hline
Ref.~\onlinecite{denis2019enhancement} FSCC+Gaunt\,\,
  & $23.37$ &  \\
Ref.~\onlinecite{prasannaa2019enhanced} QZ CCSD\,\,
  & $23.80$ &  \\
Ref.~\onlinecite{gaul2020ab} cGHF\,\,
  & $23.57$ & $20.60$ \\
Ref.~\onlinecite{gaul2020ab} cGKS\footnote{For the value of $\Omega=0.495$.}
  & $17.48$ & $15.25$ \\
\end{tabular}
\end{ruledtabular}
\end{table}

\begin{acknowledgments}
The work is supported by the Russian Science Foundation grant No. 18-12-00227.
\end{acknowledgments}
\section*{Author Declarations}

\subsection*{Conflict of interest}
The authors have no conflicts to disclose.

\section*{Availability of data}
The data that support the findings of this study are available from the corresponding author upon reasonable request.

\end{document}